\documentclass{revtex4}

\usepackage{graphicx}
\begin{document}
% You should use BibTeX and revtex.bst for references
\bibliographystyle{revtex}

% number  and your conference paper identification number on the
% title page in preprint mode. Multiple \preprint commands are allowed.
%\preprint{}

\title{Higgs bosons may be sneutrinos}
% Optional argument for running titles on pages
%\title[]{}



\author{Roland E. Allen and Antonio R. Mondragon}

\email[]{allen@tamu.edu, mondragon@tamu.edu}
%\thanks{}
\affiliation{Physics Department, Texas A\&M University, College 
Station, Texas 77843}


%\date{\today}

\begin{abstract}
In a GUT scenario with unconventional supersymmetry, Higgs bosons can be 
sneutrinos.
\end{abstract}
% \pacs{}

%\maketitle must follow title, authors, abstract and \pacs
\maketitle

%\section{}
%\label{}
%\subsection{}
%\subsubsection{}

% figures should be put into the text as floats.
% Use the graphicx package (distributed with LaTeX2e).
% See the LaTeX Graphics Companion by Michel Goosens, Sebastian Rahtz,
% \begin{figure}
% \caption{}
% \label{}
% \end{figure}

% tables follow here or maybe be put in the text
% \begin{table}
% \caption{}
% \label{}
% \end{tabular}
% \end{table}

%\begin{acknowledgments}
%\end{acknowledgments}

The next few years are full of promise for historic discoveries in 
high-energy physics. One expects that both a Higgs boson~[1-5] and
supersymmetry~[6-16] will be observed at either the Tevatron or the
LHC~[17-21], with more detailed exploration at a next-generation 
linear collider~\cite{tesla-LC-study,snowmass-LC-study}.

It was proposed long ago that Higgs fields might be superpartners of lepton
fields~\cite{fayet}, but this original proposal was ruled out because of 
two problems: (1) With standard Yukawa couplings, lepton number would 
not be conserved. (2) This proposal is incompatible with 
standard supersymmetric models for more technical reasons~\cite{Kane-review}.

Recently we proposed a very different scenario, in which both Yukawa
couplings and supersymmetry have unconventional forms~\cite{allen1,allen2}. 
The basic picture is an $SO(10)$ grand-unified gauge theory with the 
nonstandard supersymmetry described in Refs. 26 and 27.

(1) \textbf{Yukawa couplings. }Suppose that the first stage of
symmetry-breaking at the GUT scale involves a Higgs field $\phi_{GUT}$ 
which is the superpartner of the charge-conjugate of a right-handed 
neutrino field $\nu_{R}$. Then $\phi _{GUT}$ has a lepton number 
of -1 and an R-parity 
\begin{equation}
R=\left( -1\right) ^{3\left( B-L\right) +2s}=-1.
\end{equation}
This picture is compatible with the minimal scheme~\cite{collins}
\begin{equation}
SO(10) \rightarrow SU(5) \rightarrow SU(3)
\times SU(2) \times U(1) \rightarrow SU(3)\times U(1).
\end{equation}
More specifically, there are $16$ scalar boson fields, in each of $3$
generations, which are superpartners of the $16$ fermion fields per
generation in a standard $SO(10)$ theory. There are then $3$ scalar 
bosons which are superpartners associated with right-handed neutrinos, 
and $\phi _{GUT}$ is a linear combination of these $3$ bosonic fields.

Suppose also that symmetry-breaking at the electroweak scale involves a
Higgs field $\phi _{EW}$ which is the superpartner of the charge-conjugate
of a left-handed lepton field: 
\begin{equation}
\phi _{EW}=\left( 
\begin{array}{c}
\phi ^{+} \\ 
\phi ^{0}
\end{array}
\right) \qquad ,\qquad \psi _{\ell }=\left( 
\begin{array}{c}
\nu _{L} \\ 
e _{L}
\end{array}
\right) .
\end{equation}
In the simplest description, $\nu _{L}$ is the electron neutrino and 
$e _{L}$ is the left-handed field of the electron.

Finally, suppose that a typical fermion field below the GUT scale has an
effective Yukawa coupling with the form 
\begin{equation}
\lambda _{eff}=\lambda _{0}\frac{\left\langle \phi _{GUT}^{\dagger
}\right\rangle }{m_{GUT}}
\end{equation}
where $\lambda _{0}$ is dimensionless and $\left\langle \phi _{GUT}^{\dagger
}\phi _{GUT}\right\rangle \sim m_{GUT}^{2}$ with $m_{GUT}$ $\sim 10^{13}$
TeV. One then has a Dirac mass term
\begin{eqnarray}
m\psi _{L}^{\dagger }\psi _{R} &=&\psi ^{\dagger }\lambda _{eff}\left\langle
\phi _{EW}\right\rangle \psi _{R}=\lambda _{0}\psi ^{\dagger }\frac{
\left\langle \phi _{GUT}^{\dagger }\right\rangle }{m_{GUT}}\left\langle \phi
_{EW}\right\rangle \psi _{R} \\
\psi  &=&\left( 
\begin{array}{c}
\psi _{L}^{\prime } \\ 
\psi _{L}
\end{array}
\right) \quad ,\quad \left\langle \phi _{EW}\right\rangle =\left( 
\begin{array}{c}
0 \\ 
v/\sqrt{2}
\end{array}
\right) .
\end{eqnarray}
(There is another set of effective Yukawa couplings involving the
charge-conjugate Higgs fields, of course.) Since both $\phi _{GUT}$ and 
$\phi _{EW}$ have a lepton number of -1, $\phi _{GUT}^{\dagger }\phi _{EW}$
conserves lepton number. The same is obviously true of the operator in 
$v^{2}/2=\left\langle \phi _{EW}^{\dagger }\phi _{EW}\right\rangle ,$ which
determines the masses of the W bosons~\cite{cheng}. 

Since the operator in (5) is effectively dimension-four below the 
grand-unification scale, the theory is renormalizable up to this scale.
At energies above $m_{GUT}$, one has a dimension-five operator and the
theory is no longer renormalizable, but this is exactly what one expects of
a fundamental theory near the Planck scale~\cite{peskin}.

(2) \textbf{Unconventional supersymmetry.} In the present context it 
is desirable to use the same broad definition of the term ``supersymmetry'' 
that was used in Ref. 26 (in accordance with previous 
usage in various contexts, including nonrelativistic problems~[31-35]): 
A Lagrangian is supersymmetric if it is invariant under a 
transformation which converts fermions to bosons and bosons to 
fermions.

This is the fundamental meaning of supersymmetry: For every fermion there is
a bosonic superpartner and vice-versa. It is this property that gives 
credibility to supersymmetry as a real feature of nature. 
For example, it permits the cancellation of fermionic 
and bosonic radiative contributions to the mass of 
the Standard Model Higgs, which would otherwise diverge quadratically. It
also modifies the running of the $SU(3)$, $SU(2)$, and $U(1)$ coupling constants
so that they can meet at a common energy $m_{GUT}$.

In addition to this primary physical motivation for supersymmetry, there
is also a secondary and more mathematical aspect in the standard theories that
are most widely discussed -- namely, the algebra in which 
supersymmetric boson-fermion transformations of particle states 
are intimately connected to spacetime transformations of the 
inhomogeneous Lorentz group. However, there is no necessary logical 
connection between supersymmetry (as we have defined it above) and Lorentz 
invariance. Indeed, it is easily conceivable that some form of 
supersymmetry holds at the highest energies, up to the Planck 
scale, and that Lorentz invariance does not.

The supersymmetry of Refs. 26 and 27 has two unconventional aspects: First,
Lorentz invariance is not postulated, but instead automatically emerges 
in the regimes where it has been tested -- e.g., for gauge bosons and 
for fermions at energies that are far below the Planck scale. 
(The theory appears to be in agreement with the most 
sensitive experimental and observational tests of Lorentz invariance
that are currently available, largely because 
many features of this symmetry are preserved, including rotational
invariance, CPT invariance, and the same velocity $c$ for all massless
particles.) Second, gauge bosons are not fundamental, but
are instead collective excitations of the GUT Higgs field. 
In the simplest picture, the first two stages of the 
symmetry-breaking depicted in (2) can be regarded as 
follows: (i) There are three scalar boson fields $\phi _{GUT}^{i}$ which are
the superpartners of the charge-conjugates of three right-handed neutrino 
fields $\nu _{R}^{i}$. (These three generations of right-handed neutrinos
are a standard feature of $SO(10)$ grand unification~[36-42]. Through the see-saw
mechanism, they give rise to neutrino masses of about the right size to
explain recent experimental observations~[43-47].) In the first stage of
symmetry-breaking, each of the initial GUT Higgs fields acquires a vacuum
expectation value $\left\langle \phi _{GUT}^{i}\right\rangle $. At the same
time, each $\nu _{R}^{i}$ acquires a large mass~\cite{collins}. 
Then $3$ complex bosonic 
fields and $3$ fermionic fields are effectively lost. In the next stage 
depicted in (2), $24$ real bosonic fields participate in the
symmetry-breaking and are lost~\cite{collins,cheng}. Below $m_{GUT}$, 
however, one gains the initially massless vector bosons of 
the Standard Model, with $(8+3+1)\times 2=24$ bosonic degrees of 
freedom. The net result is that an equal number of bosonic and fermionic
degrees of freedom are lost, and there is still supersymmetry below $m_{GUT}$ 
down to some energy $m_{susy}\sim 1$ TeV where both bosons and fermions
acquire unequal masses.

The particular version of supersymmetry in Refs. 26 and 27 permits a pairing
of bosonic and fermionic fields like that in (3) because the initial
superpartners consist only of spin zero bosons and spin 1/2 fermions, and
because there is a relaxation of the restrictions imposed by Lorentz
invariance. We conclude that there is at least one viable scenario in which Higgs 
bosons are sneutrinos, with an R-parity of -1.

%\bibliography{your bib file}

\end{document}